\documentclass[prd, aps, nobibnotes, nofootinbib,
preprint,
showkeys,
preprintnumbers,
amsmath,
amssymb]{revtex4-1}

\usepackage{graphicx}
\usepackage{dcolumn}
\usepackage{bm}
\usepackage{hyperref}
\usepackage{mathrsfs}
\usepackage{xcolor}
\usepackage{slashed}
\usepackage{bigints}

\def\D{{\mathscr D}}

\begin{document}
\title{Emergent vortex-electron interaction from dualization}
\author{Shantonu Mukherjee}
\email{shantonumukherjee@bose.res.in}
\affiliation{S N Bose National Centre for Basic Sciences, Block JD, Sector III, Salt Lake, Kolkata 700106, India}
\author{Amitabha Lahiri}
\email{amitabha@bose.res.in}
\affiliation{S N Bose National Centre for Basic Sciences, Block JD, Sector III, Salt Lake, Kolkata 700106, India}

\date{\today}

\begin{abstract}
	
We consider the Abelian Higgs model in 3+1 dimensions with vortex lines, into which charged fermions are introduced.  This could be viewed as a model of a type-II superconductor with unpaired electrons (or holes), analogous to the boson-fermion model of high-$T_c$ superconductors but one in which the bosons and fermions interact only through the electromagnetic gauge field. We investigate the dual formulation of this model, which is in terms of a massive antisymmetric tensor gauge field $B_{\mu\nu}$ mediating the interaction of the vortex lines.  This field couples to the fermions through a nonlocal spin-gauge interaction term. We then calculate the quantum correction due to the fermions at one loop and show that due to the presence of this new nonlocal term a topological $B \wedge F$ interaction is induced in the effective action, leading to an increase in the mass of both the photon and the tensor gauge field. Additionally, we find a Coulomb potential between the electrons, but with a large dielectric constant generated by the one-loop effects.


\end{abstract}
\keywords{Boson-fermion model, vortex-fermion interaction, dual field theory, high $T_c$ superconductor.}

\maketitle

\pagebreak

\newpage
\pagenumbering{arabic}
\setcounter{page}{1}

\section{Motivation}
Vortices and vortex lines appear as solutions in the field theoretic description of many physical systems, from quantized vortices in superfluid Helium~\cite{PhysRev.136.A1194, PhysRevB.18.3197, PhysRevLett.49.1258, RevModPhys.59.533, Bewley2006VisualizationOQ} and Bose-Einstein condensates to Abrikosov lattices in type-II superconductors, all of which have been observed. They also appear in field theories which describe high energy physics, e.g. as cosmic strings which appear in many gauge theories including grand unified theories~\cite{Kibble:1976sj,Vilenkin:1984ib,Hindmarsh:1994re,Vilenkin:2000jqa}, or as color flux tubes which are conjectured to appear in non-Abelian gauge theories such as QCD, leading to color confinement~\cite{Nambu:1974zg,tHooft:1974kcl,Mandelstam:1974pi}. In this paper we will be interested in the interaction of charged fermions with vortices, which appears in the description of different physical systems.

\begin{figure}[h]
	\includegraphics[width=6cm,height=4.5cm]{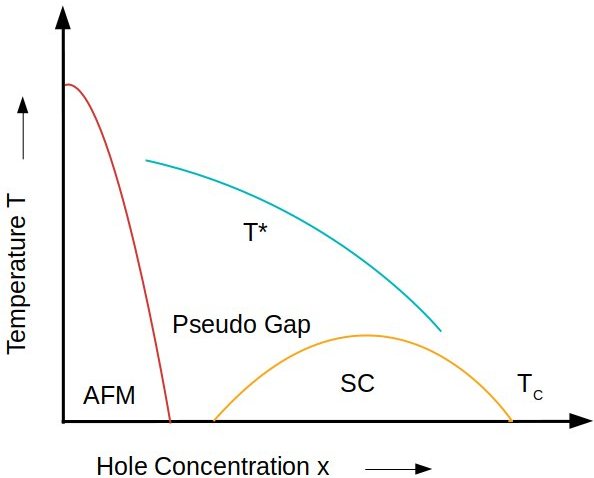}
	\centering
	\caption{Phase Diagram Of A Hole Doped High $T_c$ Superconductor}
	\label{phase-diagram}
\end{figure}

Type II superconductors allow magnetic flux to pass through in the form of lines of quantized flux~\cite{Abrikosov1957a, AbrikosovJPCS}, a phenomenon which has been experimentally confirmed~\cite{Sutton_1966, ESSMANN1967526, PhysRevLett.62.214} including in high-$T_c$ superconductors~\cite{RevModPhys.66.1125,PhysRevLett.59.2592,VINNIKOV1988421}. The phenomenon of superconductivity at high temperature has remained mysterious since its discovery in cuprates~\cite{Bednorz:1986tc}, the Bardeen-Cooper-Schrieffer (BCS) description of low-temperature superconductivity~\cite{PhysRev.108.1175, PhysRev.106.162, PhysRev.104.1189} cannot explain it.  The typical features of a high temperature superconductor (HTS) are expressed through a phase diagram~\cite{Scalapino1282,CHEN20051,Bieri} which looks like Fig.~\ref{phase-diagram}.
In the absence of doping, the cuprate material remains anti-ferromagnetic (AFM) and insulating. As doping is increased, anti-ferromagnetism does not persist and
at small doping concentration and below a temperature $T^*$ a gap opens up in the electronic energy spectrum, which is called a pseudo gap~\cite{RANDERIA19981754, doi:10.1143/JPSJ.79.044706, PhysRevB.64.064503}. As doping is increased further, superconductivity (SC) starts to appear beyond this gap.  
The presence of a  pseudo gap above $T_c$ and very small coherence length ($\sim$ 10 \AA)~\cite{PhysRevB.36.8903} indicates the formation of localized bosonic pairs of fermions (preformed pairs~\cite{CHEN20051, PhysRevB.55.3173, 2019NatSR...9.3987S}) below  $T^*$ and their condensation below $T_c$.
Based on this idea of preformed pairs, a phenomenological field-theoretic model of high temperature superconductivity was proposed by Friedberg and Lee~\cite{PhysRevB.40.6745, PhysRevB.42.4122}. In this field theory there are localized pairs, described by a bosonic field $\phi$ of charge $2e$ and mass $\sim 2m_e$, where $m_e$ is the mass of the electron. These bosons are unstable and decompose into pairs of electrons with opposite spins and these electrons recombine to form bosons. 
Thus in a large system there is always a macroscopic distribution of bosons coexisting with fermions following their respective statistical distribution laws. At temperature below $T_c$ these bosons condense, i.e. there is a large number of bosons in the zero momentum state which coexist with fermions.  This type of system with bosons and fermions coexisting in thermal equilibrium is generically referred to as boson-fermion (B-F) model.

Apart from their use in models of high-$T_c$ superconductors~\cite{Ranninger:1995, Piegari:2003, Micnas:2002, Pawlowski:2010, SALAS201637},
mixtures of bosons and fermions are studied in many other contexts as well, both experimental and theoretical. Experimental work on the properties of a mixture of Bose and Fermi gases include study of quantum degeneracy~\cite{Hadzibabic:2002, Truscott:2001, Roati:2003, Schreck:2001} and interactions~\cite{Goldwin:2004, Wu:2011, Park:2012}. Boson-fermion mixtures of dilute atomic and molecular gases at low temperatures are also studied theoretically in optical lattices to study their quantum phases including superfluid-insulator transition~\cite{Lewenstein:2004, Lewenstein:2007, Illuminati:2004, Yang:2008}. Boson-fermion systems also appear in studies of superconductor-insulator transitions~\cite{Cuoco:2004, Dubi:2007, Loh:2016}, of BCS-BEC crossover~\cite{Deng:2007,  Maska:2017}, of charged Bose liquids~\cite{Kabanov:2005}, etc. 

The boson-fermion model of Friedberg and Lee closely resembles the Abelian Higgs model~\cite{Higgs:1964pj,Higgs:1966ev,Englert:1964et,Guralnik:1964eu} as a field theory, including the appearance of vortices~\cite{PhysRevB.42.4122}. These vortices carry quantized magnetic flux, as can be derived from the minimum energy condition. The Abelian Higgs model in 3+1 dimensions contains vortex lines~\cite{Nielsen:1973cs} which are minimum energy solutions of the field equations with topologically nontrivial boundary conditions. The interaction between these lines, or strings, is mediated by a 2-form (2-index antisymmetric tensor) gauge potential $B_{\mu\nu}$ called the Kalb-Ramond field~\cite{PhysRevD.9.2273}. In the symmetry broken phase and when vortex strings are present, the Abelian Higgs model can be written in terms of the string world sheet and the 2-form gauge field $B_{\mu\nu}$ using dualization~\cite{PhysRevD.48.2493,Chatterjee:2006iq, Ramos:2005yy, Franz_2007,2011NJPh...13c3004B}. Our goal for this work is to do this with fermions, i.e., to dualize the boson-fermion system in presence of vortices and reach what should be a useful starting point for the interaction of vortex lines with unpaired charged fermions. 

Of particular interest is the coupling between the 2-form gauge field $B_{\mu\nu}$ and fermions. In an earlier work it was proposed that the 2-form field couples nonlocally to a topologically conserved current of the electrons~\cite{Choudhury:2015rua}, 
\begin{equation}
\int d^4x B_{\mu\nu} \frac{1}{\square} J^{\mu\nu},
\end{equation} 
where $J^{\mu\nu} = \epsilon^{\mu\nu\rho\lambda}\partial_\rho (\bar{\psi}\gamma_\lambda \psi)$ is the 4-dimensional ``curl" of the conserved fermion current. The nonlocal current $\dfrac{1}{\square}J^{\mu\nu}$\,, more specifically its $\{0i\}$ component, 
contains the spin magnetic moment density as a contribution from the spin part. Then we can say that the coupling $B_{0i}\dfrac{1}{\square}J^{0i}$  corresponds to a spin-spin interaction mediated via the 2-form gauge field $B$. If we find this interaction here, we will be able to say that we have found a local theory, namely that of the boson-fermion mixture, which has a description containing this ``spin-gauge interaction''. The electrons interact via photons as well, and quantum corrections due to fermion loops give rise to an effective $B\wedge F$ interaction. This term is central to the topological mass mechanism in 3+1 dimensions~\cite{Allen:1990gb}, analogous to the Chern-Simons term in 2+1 dimensions~\cite{Schonfeld:1980kb,Deser:1982vy,Deser:1981wh} which can also be generated by fermion loops~\cite{Dorey:1991kp}. We will be working with the Abelian Higgs model in the broken phase, in which the photon is already massive. Thus we expect that the mass of the photon will only be modified in this case.
However, the $B\wedge F$ interaction and its lower dimensional version, the mixed Chern-Simons action, have also been useful in theories of topological superconductors, topological insulators, and quantum Hall effect~\cite{CHO20111515, Diamantini:2014iqa, PhysRevB.90.235118, PhysRevD.89.107702, PhysRevLett.112.016404, PhysRevX.5.021029, Hirono2019}. Our results should be useful for these systems. 

The outline of our paper is as follows. In Sec.~\ref{dual} we dualize the Abelian Higgs model in the presence of vortex lines (strings) and charged fermions which couple through electromagnetic interactions, culminating in a nonlocal dual Lagrangian involving strings and the 2-form field which mediates interstring interactions. In Sec.~\ref{induced} we derive the effective action by taking into account 1-loop corrections due to fermion loops. This generates a $B\wedge F$ interaction, which affects the propagators of both $B_{\mu\nu}$ and $A_\mu$. Then in Sec.~\ref{force} we calculate the static potential between nonrelativistic fermions taking into account all the interactions as well as the 1-loop correction and end with some comments.

\section{Dual Lagrangian for Boson-Fermion system}\label{dual}
We first determine the dual of the field theory describing a boson-fermion system in the presence of vortices. Even though all particles in this system move non-relativistically, we will work with a four dimensional relativistic field theory. This is because the field theoretic duality we consider is most conveniently constructed in four dimensions and for relativistic theories, and also because we will be able to use several standard results from usual quantum electrodynamics.

We start with the Abelian Higgs model where the gauge field $A_\mu $ is minimally coupled to unpaired charged fermions in addition to the complex scalar Higgs field. The Lagrangian of our system is thus
\begin{equation}
    \mathscr{L} = -\frac{1}{4}F_{\mu\nu}F^{\mu\nu} + \frac{1}{2}D^\mu \phi\textsuperscript{\textdagger}D_\mu \phi + \bar{\psi}(i\gamma^\mu \partial_\mu - m)\psi
-V(\phi\textsuperscript{\textdagger}\phi)
-e A_\mu \bar{\psi}\gamma^\mu\psi\,,
\end{equation}
where $\phi$ is a complex scalar field of charge $q$ ($2e$ if $\phi$ describes Cooper pairs), $\psi$ is the fermionic field with charge $e$\,, $D_\mu \phi = \partial_\mu+iqA_\mu\phi\,$ and $V(\phi^\dagger \phi)  = \dfrac{\lambda}{4}(\phi^\dagger \phi - v^2)^2\,$ is the symmetry breaking potential.

As is well known, topologically stable structures like vortices or flux tubes can appear in this theory because the circle on which $\phi$ lies at its minimum is mapped on a circle at infinity. If there is a vortex in a plane, corresponding to a flux tube cutting through the plane as we will consider, the phase of $\phi$ becomes multivalued as we go around a circle at infinity. The vacuum condition $D_\mu \phi = 0\,$ then leads to quantization of magnetic flux in the vortex,
\begin{equation}
       \oint\limits_C A_\mu dx^\mu = -\frac{2n\pi}{q}\,,
       \label{flux-quantized}
\end{equation}
where $C$ is a circle at infinity and $n$ is the winding number, i.e., the number of times the phase of $\phi$ winds around the vortex. It is a topological quantum number and describes the quantization of topological charge, while $\dfrac{2\pi}{q}$ is the quantum of magnetic flux passing through the vortex. To consider vortices explicitly we express $\phi$ in polar form,
\begin{equation}
    \phi = v f \exp({i\chi})\,.
\end{equation}
The function $f$ vanishes along the core of the flux tube and reaches $f = 1$ far from the core region\footnote{The zero of $f$ on a plane is the location of a vortex. A locus of zeroes in space defines a vortex string or a flux tube. }. The Lagrangian, including a gauge-fixing term, then takes the form
\begin{align}
{\mathscr L} =&  -\frac{1}{4}F_{\mu\nu}F^{\mu\nu} 
+ \frac{1}{2} v^2\partial_\mu f \partial^\mu f + \frac{1}{2}v^2 f^2(\partial_\mu \chi + q A_\mu)(\partial^\mu \chi +q A^\mu) \notag \\
&\qquad -\frac{1}{2\xi}(\partial_\mu A^\mu)^2 
- \frac{\lambda}{4}(f^2 - 1)^2 + \bar{\psi}(i\gamma^\mu \partial_\mu - m)\psi - eA_\mu \bar{\psi}\gamma^\mu\psi\,.
\end{align}

We shall dualize the theory, starting from the partition function 
\begin{equation}
\mathcal{Z} = \int {\D}A_\mu {\D}f {\D}\chi {\D}\bar{\psi} {\D}\psi \exp\left({i\int d^4x\, \mathscr{L} }\right)\,.
\end{equation}
The duality transformation takes us from the Higgs picture above, where the degrees of freedom are adequately described by a a charged Higgs minimally coupled to the electromagnetic gauge field, to an equivalent vortex picture in which vortices interact through the second rank antisymmetric tensor Kalb-Ramond field. To implement this we first linearize the term $\dfrac{1}{2}v^2 f^2(\partial_\mu \chi + q A_\mu)^2 $ by introducing an auxiliary field through a Gaussian integral into the partition function as
\begin{equation}
  N  \int {\D}C_\mu\exp{\left(-i\int d^4x\left[\frac{C_\mu}{\sqrt{2}v} + \frac{v}{\sqrt{2}}f(\partial_\mu \chi + qA_\mu)\right]^2\right)} = 1\,.
\end{equation}
Then we can write the partition function as 
\begin{align}
 \mathcal{Z} =& \int {\D}A_\mu {\D}f {\D}\chi {\D}\bar{\psi} {\D}\psi {\D}C_\mu \notag\\
    &\qquad  \exp\bigg(i\int d^4x \bigg( -\frac{1}{4}F_{\mu\nu}F^{\mu\nu} + \frac{1}{2} v^2\partial_\mu f \partial^\mu f -\frac{C_\mu C^\mu}{2v^2} - C^\mu f (\partial_\mu \chi + qA_\mu)\notag \\
& \qquad \qquad \qquad \qquad - \frac{1}{2\xi}(\partial_\mu A^\mu)^2 
- V(f^2) + \bar{\psi}(i\gamma^\mu \partial_\mu - m)\psi - eA_\mu \bar{\psi}\gamma^\mu\psi \bigg)\bigg)\,.
\end{align}
As mentioned earlier, the phase $\chi$ is multivalued around a vortex string and the value of $\chi$ changes by $2n\pi$  as one goes around the vortex string $n$ times, $n$ being the winding number. So we decompose $\chi = \chi^r + \chi^s$, where the superscript $s$ indicates the singular part of $\chi$ describing a vortex configuration and $r$ denotes the regular part which is single valued and corresponds to fluctuations around a given vortex configuration. 
By doing an integration by parts on the term $C^\mu f \partial_\mu \chi^r$ we can shift the partial derivative onto $C^\mu f$ and then integrate over $\chi^r$ producing a delta function. Thus we can write 
\begin{align}
 \mathcal{Z} = \int &{\D}A_\mu {\D}f {\D}\chi^s {\D}\bar{\psi} {\D}\psi {\D}C_\mu  \delta(\partial_\mu(C^\mu f)) \notag\\
&     \exp\left(i\int d^4x \left( -\frac{1}{4}F_{\mu\nu}F^{\mu\nu} + \frac{1}{2} v^2\partial_\mu f \partial^\mu f -\frac{C_\mu C^\mu}{2v^2} - C^\mu f (\partial_\mu \chi^s + q A_\mu)
\right.\right. \notag\\
 & \qquad \quad \left.\left. - \frac{1}{2\xi}(\partial_\mu A^\mu)^2 
 - V(f^2) + \bar{\psi}(i\gamma^\mu \partial_\mu - m)\psi - eA_\mu \bar{\psi}\gamma^\mu\psi \right)\right)\,.
\end{align}
%
The delta function can be solved by introducing an antisymmetric tensor (2-form) potential 
$B_{\mu\nu}$ and setting $ C^\mu = \dfrac{1}{2 f}\varepsilon ^{\mu\nu\rho\lambda} \partial_\nu B_{\rho\lambda}\,,$ which allows us to write the partition function as 
%
\begin{align}
 \mathcal{Z} = \int &{\D}A_\mu {\D}f {\D}\chi^s {\D}\bar{\psi} {\D}\psi{\D}B_{\mu\nu} \notag\\
&     \exp\bigg(i\int d^4x \bigg( -\frac{1}{4}F_{\mu\nu}F^{\mu\nu} + \frac{1}{2} v^2\partial_\mu f \partial^\mu f +\frac{1}{12v^2f^2}H^{\nu\rho\lambda} H_{\nu\rho\lambda}  
- \frac{1}{2\xi}(\partial_\mu A^\mu)^2 
\notag\\ 
&-\frac{1}{2}\varepsilon^{\mu\nu\rho\lambda} B_{\rho\lambda} \partial_\mu\partial_\nu \chi^s  -\frac{q}{2}\varepsilon^{\mu\nu\rho\lambda} \partial_\nu B_{\rho\lambda} A_\mu
- V(f^2) + \bar{\psi}(i\gamma^\mu \partial_\mu - m)\psi - eA_\mu \bar{\psi}\gamma^\mu\psi \bigg)\bigg).
\end{align}
Here we have defined $H_{\nu\rho\lambda} = \partial_\nu B_{\rho\lambda}+\partial_\rho B_{\lambda\nu}+\partial_\lambda B_{\nu\rho}$ as the field strength of the 2-form field. 
The curl of the velocity of the scalar field (or supercurrent)  is called vorticity,  $\Sigma^{\rho\lambda}=\varepsilon ^{\mu\nu\rho\lambda} \partial_\mu\partial_\nu \chi^s $. 
Around a vortex this quantity is non-zero and in case of a straight rod like array of vortices, i.e. a vortex line or a flux tube, along the $Z$-axis, it is given by
\begin{equation}
(\partial_x\partial_y - \partial_y\partial_x)\chi^s\\
= 2n\pi \delta^2(\vec{\rho}).
\end{equation}  
This expresses the location of the vortices in the $X-Y$ plane.
By dualizing it we get the world sheet of the vortex line in 3+1 dimensions,
\begin{align}
    \Sigma^{\rho\lambda}& = \varepsilon ^{\mu\nu\rho\lambda} \partial_\mu\partial_\nu \chi^s \notag\\
 & = \int d\sigma_{\mu\nu}\delta(x-X)\,.
\end{align}
$X^\mu$ are the coordinates of the world sheet of the vortex line and  $d\sigma_{\mu\nu}=d\tau ds \dfrac{\partial(X_\mu,X_\nu)}{\partial(s,\tau)}$ is the surface element over the world sheet. 
Thus we write the partition function as
\begin{align}
 \mathcal{Z}  = \int & {\D}A_\mu {\D}f {\D}\chi^s {\D}\bar{\psi} {\D}\psi {\D}B_{\mu\nu} \notag\\
&  \exp\left(i\int d^4x \left( -\frac{1}{4}F_{\mu\nu}F^{\mu\nu} + \frac{1}{2} v^2\partial_\mu f \partial^\mu f +\frac{1}{12v^2f^2}H^{\nu\rho\lambda} H_{\nu\rho\lambda}
-\frac{1}{2} B_{\rho\lambda}\Sigma^{\rho\lambda} 
\right.\right.  \notag\\
 & \qquad \left.\left.
 -\frac{q}{2}\varepsilon ^{\mu\nu\rho\lambda} \partial_\nu B_{\rho\lambda} A_\mu
 - \frac{1}{2\xi}(\partial_\mu A^\mu)^2 
 - V(f^2) + \bar{\psi}(i\gamma^\mu \partial_\mu - m)\psi - eA_\mu \bar{\psi}\gamma^\mu\psi \right)\right)\,. 
\end{align}
Next we rename the currents $\frac{q}{2}\varepsilon ^{\mu\nu\rho\lambda} \partial_\nu B_{\rho\lambda} = J^\mu_H$ and $e \bar{\psi}\gamma^\mu\psi= J_\psi^\mu $\,, separate out the terms which depend on $A_\mu$ from the rest of the partition function and then integrate over $A_\mu$\,,
\begin{align}
\int {\D}A_\mu \exp&\left(i\int d^4x \left(-\frac{1}{4}F_{\mu\nu}F^{\mu\nu} - \frac{1}{2\xi}(\partial_\mu A^\mu)^2 + A_\mu(J^\mu_H + J^\mu_\psi) \right)\right) \notag\\
& = 
{\mathscr N}
\exp\bigg(-\frac{i}{2}\int d^4x d^4y (J^\mu_H + J^\mu_\psi) \Delta_{\mu\nu} (J^\nu_H + J^\nu_\psi)\bigg)\,.
\label{PI.separate}
\end{align}
Here $\Delta_{\mu\nu}$ is the Green function corresponding to the operator
\begin{equation}
\Delta^{-1}_{\mu\nu} = g^{\mu\nu}\Box - (1-\frac{1}{\xi})\partial^\mu\partial^\nu \,.
\end{equation}
In momentum space it is given by
\begin{equation}
\Delta_{\mu\nu}(k)= - \frac{g^{\mu\nu}-(1-\xi)\frac{k^\mu k^\nu}{k^2}}{k^2 + i\epsilon}\,.
\end{equation}
We will suppress the $+i\epsilon$ in what follows, but it is present in each of the propagators appearing below.
The integration over $A_\mu$ has produced a normalization factor ${\mathscr N}$ which does not contribute to the rest of the partition function. 
We can thus write the action as 
\begin{align}
S =  \int d^4x & \bigg(\frac{1}{2} v^2\partial_\mu f \partial^\mu f +\frac{1}{12v^2f^2}H^{\nu\rho\lambda} H_{\nu\rho\lambda}
-\frac{1}{2} B_{\rho\lambda}\Sigma^{\rho\lambda}  
 - V(f^2) + \bar{\psi}(i\gamma^\mu \partial_\mu - m)\psi\bigg) \notag\\ 
 & \qquad -\frac{1}{2}\int d^4x d^4y \left(J^\mu_H + J^\mu_\psi\right)(x) \Delta_{\mu\nu}(x,y) \left(J^\nu_H + J^\nu_\psi\right)(y)\,.
\end{align} 
We can further simplify the last term. Note that since $J^\mu_H$ is a (topologically) conserved current, the second term in $\Delta_{\mu\nu}$ annihilates it, so we can write 
\begin{equation}
\frac{1}{2}\int d^4x d^4y J^\mu_H(x)\Delta_{\mu\nu}(x,y)J^\nu_H(y) = - \int d^4x \frac{q^2}{12}H_{\nu\rho\lambda}\frac{1}{\square}H^{\nu\rho\lambda}\,,
\end{equation}
as well as 
\begin{equation}
\int d^4x d^4y\; J^\mu_H(x)\Delta_{\mu\nu}(x,y)J^\nu_\psi(y) =\int d^4x \;\frac{1}{2}e q B^{\mu\nu}\varepsilon_{\mu\nu\rho\lambda} \partial^{\rho}\frac{1}{\square}\bar{\psi}\gamma^{\lambda}\psi\,.
\label{nonlocal}
\end{equation}
In order to understand the remaining part, which is quadratic in the fermion current 
$J^\mu_\psi$\,, we note that it is exactly what we would get if we integrate over $A_\mu$
ordinary quantum electrodynamics, i.e.
\begin{align}\label{PI.QED}
\int {\D}A_\mu \exp\left(i\int d^4x \left(-\frac{1}{4}F_{\mu\nu}F^{\mu\nu} +A_\mu J^\mu_\psi + \frac{1}{2\xi}(\partial_\mu A^\mu)^2 \right)\right) \notag\\ 
= {\mathscr N}_0 \exp\left(-\frac{i}{2}\int d^4x d^4y\; J^\mu_\psi(x)\Delta_{\mu\nu}(x,y)J^\nu_\psi(y)\right)\,,
\end{align}
where ${\mathscr N}_0$ is a normalization factor. 

Thus after collecting all these terms, we can write the dual Lagrangian as
\begin{equation}\label{dual-lagrangian}
\boxed{\begin{aligned}
   \mathscr{L}  =  & -\frac{1}{4}F_{\mu\nu}F^{\mu\nu}+ \bar{\psi}(i\gamma^\mu \partial_\mu - m)\psi -eA_\mu\bar{\psi}\gamma^{\mu}\psi -\frac{1}{2}e q B^{\mu\nu}\varepsilon_{\mu\nu\rho\lambda} \partial^{\rho}\frac{1}{\square}\bar{\psi}\gamma^{\lambda}\psi +\frac{1}{2} v^2\partial_\mu f \partial^\mu f \\ &
     \qquad \qquad + \frac{1}{12 v^2}H^{\nu\rho\lambda}\left(\frac{1}{f^2} + \frac{q^2 v^2}{\square}\right) H_{\nu\rho\lambda} 
   -\frac{1}{2} B_{\rho\lambda}\Sigma^{\rho\lambda} - V(f^2)\,,
\end{aligned}}
\end{equation}
where we have suppressed gauge-fixing terms for $A_\mu$\, or $B_{\mu\nu}$\,.
Thus starting from a system containing vortex strings and described by an Abelian Higgs model in the broken phase in which charged fermions are also present, we have arrived at the dual Lagrangian of Eq.~(\ref{dual-lagrangian}) in which the Kalb-Ramond field $B_{\mu\nu}$~\cite{PhysRevD.9.2273} couples to a topologically conserved nonlocal tensor current,
\begin{equation}\label{Jmunu}
J^{\mu\nu} = 
\frac{1}{2}  q \varepsilon^{\mu\nu\rho\lambda}  {\square}^{-1}  \partial_{\rho}J_\lambda\,,
\end{equation}
with $J^\mu$ being the conserved electron current. The  conserved charge density $J^{0i}$ for this current can be split into orbital and spin parts, with the spin contribution for nonrelativistic electrons being the intrinsic spin density of the electron,
\begin{equation}\label{spin_NR}
\left(J^{0i}_{\rm spin}\right)_{\rm NR} \propto \psi^\dagger \sigma^i \psi\,,
\end{equation}
up to dimensionful constants, when the charge is time-independent and cannot accumulate. Thus in other words we have a gauge theory in which the gauge potential mediating string-string interaction couples to the spin current of charged fermions~\cite{Choudhury:2015rua}.

The Lagrangian of Eq.~(\ref{dual-lagrangian}) is invariant, not only with respect to the usual gauge transformation 
$A_\mu \to A_\mu + \partial_\mu \lambda$ for arbitrary real functions $\lambda$\,, but also 
under the vector (or extended or higher or Kalb-Ramond) gauge transformation 
$B_{\mu\nu} \rightarrow B_{\mu\nu} + \partial_\mu \Lambda_\nu -  \partial_\mu \Lambda_\nu$\,,
provided 
\begin{equation}
    \partial_\mu \Sigma^{\mu\nu} = 0\,.
\end{equation}
This shows that vortex lines must either form closed loops or be infinitely long (or end at the boundaries of the superconducting region) as the world-sheet current is conserved by itself. This is of course expected as magnetic field lines must either close on themselves or go out to infinity, since there are no magnetic monopoles.

\section{Induced $B\wedge F$ Term}\label{induced}
In order to see the effect of the nonlocal coupling on the boson-fermion system, let us calculate the quantum corrections at one fermion loop. We will do this by first setting $f\rightarrow 1$\,, which corresponds to the limit of the flux tubes being very thin. We also redefine  $\frac{1}{v} B_{\mu\nu}$ as $B_{\mu\nu}$ for convenience of calculations. 

The partition function then becomes
\begin{align}
   \mathcal{Z} = \int &{\D}\chi^s {\D}\bar{\psi} {\D}\psi {\D}B_{\mu\nu}{\D}A_\mu \notag\\
&  \exp\bigg(i\int d^4x \bigg(  -\frac{1}{4}F_{\mu\nu}F^{\mu\nu}+ \bar{\psi}(i\gamma^\mu \partial_\mu - m)\psi -eA_\mu\bar{\psi}\gamma^{\mu}\psi \notag\\ & 
 -\frac{1}{2}e MB^{\mu\nu}\varepsilon_{\mu\nu\rho\lambda} \partial^{\rho}\frac{1}{\square}\bar{\psi}\gamma^{\lambda}\psi 
 + \frac{1}{12}H_{\nu\rho\lambda}(1+ {M^2}\,{\square}^{-1})H^{\nu\rho\lambda} 
 -\frac{v}{2} B_{\rho\lambda}\Sigma^{\rho\lambda}  \bigg),
\end{align}
where we have written $M=qv\,.$ The nonlocal term involving $M^2\,\Box^{-1}$ is a mass term for $B_{\mu\nu}$ and is sometimes called a Meissner term for that reason~\cite{2017PRB96p5115B, Beekman2011_FOP}\,.

The nonlocal interaction term between the $B_{\mu\nu}$-field and the fermion can be written as 
\begin{equation}
\int d^4x\; \frac{1}{2}e MB^{\mu\nu}\varepsilon_{\mu\nu\rho\lambda} \partial^{\rho}\frac{1}{\square}\bar{\psi}\gamma^{\lambda}\psi = \int d^4x\; \frac{1}{2}e M\varepsilon_{\mu\nu\rho\lambda} \bar{\psi}\gamma^{\mu}\psi\frac{1}{\square}\partial^{\nu}B^{\rho\lambda}\,,
\end{equation}
as can be seen directly from where it first appeared in Eq.~(\ref{nonlocal}).
For convenience of calculations let us now define an ``effective gauge field'' $A_\mu^{\rm{eff}}$ as
$A_\mu^{\rm{eff}} = A_\mu +{M}{\square}^{-1}F_\mu$  where we have written $ F_\mu = \frac{1}{2}\varepsilon_{\mu\nu\rho\lambda}\partial^{\nu}B^{\rho\lambda} $\,. 
Quantum corrections to the action due to fermion loops are calculated in the standard textbook method~\cite{Peskin:1995ev},\cite{Ramond:1981pw}: We expand $\psi = \psi_0 + \eta$\,, where $\psi_0$ is a solution of the equation of motion $\dfrac{\delta S}{\delta\bar{\psi}}\Big\vert_{\psi = \psi_0} = 0\,,$ similarly for $\bar{\psi}_0$\,, then integrate over $\eta\,, \bar{\eta}$\, to first order in $e^2$\, for one loop,
\begin{align}
     \int {\D}\bar{\eta} {\D}\eta\; &\exp\left(i\int d^4x\; \bar{\eta}\left(i\gamma^\mu \partial_\mu - m-e \gamma^\mu A_\mu^{\rm{eff}}\right)\eta \right)
    \notag\\     & 
      \sim \exp\left[- \frac{i}{2}\int\frac{d^4k}{(2\pi)^4} \Pi(k^2) A_\mu^{\rm{eff}}(-k) \left(g^{\mu\nu} k^2- k^\mu k^\nu\right) A_\nu^{\rm{eff}}(k)\right]\,.
\end{align}
Here $ A_\mu^{\rm{eff}}$ is defined as before, while $\Pi(p^2)$ includes the effect of modes up to a cutoff $\Lambda$ and is given by
 \begin{equation}\label{1-loop}
     \Pi(k^2) = \frac{e^2}{2\pi^2}\int_0^1 z(1-z)\left[ \ln\left(1 + \frac{\Lambda^2}{m^2-k^2 z(1-z)}\right) - \frac{\Lambda^2}{\Lambda^2 + m^2-k^2 z(1-z)}\right],                         
\end{equation}
for a cutoff $\Lambda$. The natural cutoff scale in the system of vortices and electrons is the thickness of the vortex string.  This is typically comparable to the atomic scale, so we set $\Lambda\ll m$  (also $|k^2|\ll m^2$\,) to find at the leading order
\begin{equation}\label{1-loop-Pi}
     \Pi(k^2) = \frac{e^2}{24\pi^2} \frac{\Lambda^4}{m^4} + \cdots\,,
 \end{equation}
ignoring terms of the order of $\frac{\Lambda^6}{m^6}\,, \frac{\Lambda^4 k^2}{m^2}\,, \frac{\Lambda^2 k^4}{m^6}$\,.
We can now write the Lagrangian including the loop correction as 
\begin{align}
{\mathscr L} &=   -\frac{1}{4}F_{\mu\nu}F^{\mu\nu} 
 + \frac{1}{12}H_{\nu\rho\lambda}(1+ {M^2}\,{\square}^{-1})H^{\nu\rho\lambda} 
-\frac{v}{2} B_{\rho\lambda}\Sigma^{\rho\lambda}\, 
 - \frac{e^2}{24\pi^2} \frac{\Lambda^4}{m^4} \left(\frac{1}{4} F^{\rm{eff}}_{\mu\nu}F^{\rm{eff}\mu\nu}\right)\,,
\end{align}
where we have suppressed terms involving $\psi_0$\,.
Recalling the definition of $A_\mu^{\rm{eff}}$\,, we can write
\begin{equation}
F^{\rm{eff}}_{\mu\nu}F^{\rm{eff}\mu\nu} = F_{\mu\nu} F^{\mu\nu} - M \epsilon_{\mu\nu\rho\lambda}F^{\mu\nu} B^{\rho\lambda} + \frac{1}{3}M^2 H_{\mu\nu\lambda}\frac{1}{\square}H^{\mu\nu\lambda}\,,
\end{equation}
after taking into account an integration by parts.
Writing $Z=  \dfrac{e^2}{24\pi^2} \dfrac{\Lambda^4}{m^4}\,,$ we rewrite the Lagrangian as
\begin{align}
\label{L.renorm}
{\mathscr L} 
  = & -\frac{1}{4}\left(1+Z\right)F_{\mu\nu}F^{\mu\nu} +\frac{1}{12}H^{\nu\rho\lambda}\left(1 + \left(1-Z\right)\frac{M^2}{\Box}\right) H_{\nu\rho\lambda}
+  \frac{1}{4}Z M \epsilon_{\mu\nu\rho\lambda}F^{\mu\nu} B^{\rho\lambda}  \notag \\
& \quad -\frac{v}{2} B_{\rho\lambda}\Sigma^{\rho\lambda}  
 +\bar{\psi_0}(i\gamma^\mu \partial_\mu - m)\psi_0  
 -eA_\mu\bar{\psi_0}\gamma^{\mu}\psi_0 
 -\frac{1}{2}e MB^{\mu\nu}\varepsilon_{\mu\nu\rho\lambda} \partial^{\rho}\frac{1}{\square}\bar{\psi_0}\gamma^{\lambda}\psi_0 \,.
\end{align}
If we now rescale $A_\mu \rightarrow \sqrt{1+Z} A_\mu$ and also define the ``renormalized charge'' $e^2_R = e^2\left(1 + Z\right)^{-1}\simeq {e^2}{\left(1 - Z\right)} $\,, 
we obtain the Lagrangian in the form
\begin{align}
\mathscr{L} = & -\frac{1}{4}F_{\mu\nu}F^{\mu\nu}  
+\frac{1}{12}H^{\nu\rho\lambda} \left(1 + \frac{M_R^2}{\square}\right) H_{\nu\rho\lambda} +\frac{1}{4}  Z M_R \epsilon_{\mu\nu\rho\lambda}F^{\mu\nu} B^{\rho\lambda}
  -\frac{v}{2} B_{\rho\lambda}\Sigma^{\rho\lambda} \notag \\ 
  & \qquad +\bar{\psi_0}(i\gamma^\mu \partial_\mu - m)\psi_0 -e_R A_\mu\bar{\psi_0}\gamma^{\mu}\psi_0 
 -\frac{1}{2}e_R M_R B^{\mu\nu}\varepsilon_{\mu\nu\rho\lambda} \partial^{\rho}\frac{1}{\square}\bar{\psi_0}\gamma^{\lambda}\psi_0 \,.
\label{Lag-full.1-loop}
\end{align}
Here we have written $M_R^2= M^2\left(1-Z\right)\simeq \frac{q^2v^2}{\left(1+Z\right)}$ as the ``renormalized mass'' of the gauge boson. Note that we can write $M_R = {q_R v}$ since all electric charges should be renormalized the same way, so that $q_R = \frac{q}{\sqrt{1+Z}}\,.$ There is also an induced $B\wedge F$ term with a coefficient which depends on the cutoff $\Lambda$. 

The coefficient of the induced $B\wedge F$ term is very small and depends on the cutoff, which is in turn determined by the properties of the system.
As we shall see now, this term will increase the mass of the gauge fields $A_\mu$ or $B_{\mu\nu}$\,. By the mass of the gauge fields we mean the pole of the propagator of the relevant field, which can be calculated either by summing an infinite series or by taking only the part of the Lagrangian quadratic in the fields and eliminating one gauge field in favor of the other. Let us use the second method to find the poles in the propagators, starting with $B_{\mu\nu}$\,.
We separate out the quadratic terms containing $A_\mu$\,,
\begin{align}
\mathcal{Z}_A
& = \int \mathscr{D}A_{\mu} \exp\left(i\int d^4x \left(\frac{1}{2}A_\mu K^{\mu\nu}A_\nu+\frac{1}{4}  Z M_R \epsilon_{\mu\nu\rho\lambda}F^{\mu\nu} B^{\rho\lambda}\right)\right) \,,
\end{align}
where $K_{\mu\nu}$ is the invertible operator $g_{\mu\nu}\square -(1-\frac{1}{\xi}) \partial_\mu \partial_\nu \,.$
We complete the square and write
\begin{equation}
\mathcal{Z_A} = \left(\int \mathscr{D}A^\prime_{\mu}\exp i\int d^4x\, \frac{1}{2}A^\prime_\mu K^{\mu\nu} A^\prime_\nu \right) \exp\left(i\int d^4x \left(\frac{Z^2 M_R^2}{12}H^{\sigma\rho\lambda} \frac{1}{\Box}  H_{\sigma\rho\lambda} 
\right)\right).
\end{equation}
The integration over $A'_\mu$ provides a normalization factor while the second term gets added to the Lagrangian. The mass term of $B_{\mu\nu}$ becomes
$\frac{1}{12}(1+Z^2) M_R^2 H^{\nu\rho\lambda}\, {\square}^{-1} H_{\nu\rho\lambda}$\,, resulting in a shift of the coefficient of the Meissner term so that  $M_B = M_R\sqrt{1+Z^2}$ is the new mass of $B$\,. Thus the mass of the interstring gauge potential $B_{\mu\nu}$ increases because of quantum effects due to fermion loops. 

Let us now see what happens to the propagator of $A_\mu$ if we integrate out $B_{\mu\nu}$ instead. We start from 
\begin{equation}
\mathcal{Z}_B
 = \int {\D}B_{\mu\nu} \exp\bigg(-i\int d^4x \left(\frac{1}{4}B_{\mu\nu}M^{\mu\nu\rho\lambda}B_{\rho\lambda}
 - \frac{1}{4} Z M_R \epsilon^{\mu\nu\rho\lambda} B_{\mu\nu}  F_{\rho\lambda}
\right)\bigg)\,,
\end{equation}
where for convenience we have written  
\begin{equation}\label{B.kinetic}
M^{\mu\nu\rho\lambda} = \left(\Box  + M_R^2 \right) g^{\mu[\rho}g^{\lambda]\nu}
+ \left(1 + M_R^2 \Box^{-1} - \frac{1}{\eta} \right)\left(g^{\nu[\rho}g^{\lambda]\sigma }\partial_\sigma \partial^\mu - g^{\mu[\rho}g^{\lambda]\sigma }\partial_\sigma \partial^\nu\right)\,.
\end{equation} 
We can now perform the integral by completing the square and find 
\begin{equation}
\mathcal{Z}_B = N^\prime \exp\left(-\frac{i}{4}\int d^4x d^4y\, Z^2 M_R^2 F_{\mu\nu} \frac{1}{\square+ M_R^2}F^{\mu\nu} \right)\,.
\end{equation}
This is added to the action for $A_\mu$\,,  so that the quadratic term in the Lagrangian becomes
\begin{equation}
    -\frac{1}{4}F_{\mu\nu}\left(1 + \frac{Z^2 M_R^2}{\square+ M_R^2}\right) F^{\mu\nu}-\frac{1}{2\xi}(\partial_\mu A^\mu)^2 \,.
\end{equation}
A similar expression was found in three dimensions in the closely related context of the disorder field~\cite{Kleinert:2008zzb} in a superconductor, but in the absence of charged fermions. The conclusion there was that at high temperatures, vortex lines proliferate and Meissner effect is destroyed, leaving a long range Coulomb-like interaction with a complicated dispersion relation. 

In our case, charged fermions are present and Meissner effect has not been destroyed. The propagator of the field $A_\mu$ is 
\begin{equation}
    G_{\mu\nu} = -\left(\frac{g_{\mu\nu}}{p^2}\frac{p^2-M_R^2}{p^2-M_R^2(1+Z^2)} + \frac{p_\mu p_\nu}{p^4} \frac{p^2(\xi-1)- M_R^2 \left(\xi(Z^2+1)-1\right)}{p^2-M_R^2(1+Z^2)}\right)\,.
    \label{Green-A}
\end{equation}
The second term in the Green function will disappear when a conserved current couples to it, while the first term can be written as
\begin{equation}\label{photon-propagator}
G_{\mu\nu} = -\left(\frac{1}{1+Z^2}\frac{g_{\mu\nu}}{p^2} + \frac{Z^2}{1+Z^2}\frac{g_{\mu\nu}}{p^2-M_R^2(1+Z^2)}\right)\,.
\end{equation}
Unlike the usual mechanism of topological mass generation using a $B\wedge F$ interaction where the massive $B_{\mu\nu}$ field is `dual' to the massive $A_\mu$ field in the sense that the propagating degrees of freedom can be described equally well by either field, the boson-fermion system shows a distinction between the two alternative descriptions in presence of vortices.

\section{Force Law}\label{force}
Another window to the physics of the system is provided by the force law as experienced by the charged fermions, which we proceed to derive now. We will find it by integrating out both the gauge fields $A_\mu$ and $B_{\mu\nu}$ and then taking the nonrelativistic limit for the fermionic currents. We start by integrating out $B_{\mu\nu}$ from the Lagrangian of Eq.~(\ref{Lag-full.1-loop}), which includes corrections up to one fermion loop,
\begin{equation}
\mathcal{Z}_B
 = \int {\D}B_{\mu\nu} \exp\bigg(-i\int d^4x \left(\frac{1}{4}B_{\mu\nu}M^{\mu\nu\rho\lambda}B_{\rho\lambda}
- \frac{1}{2} B_{\mu\nu}J^{\mu\nu}\right)\bigg)\,,
\end{equation}
where we have written $J_{\mu\nu} = \frac{1}{2}ZM_R\epsilon_{\mu\nu\rho\lambda}F^{\rho\lambda} - v\Sigma_{\mu\nu}- e_RM_R\epsilon_{\mu\nu\rho\lambda}\partial^\rho (\square^{-1})J^\lambda\,,$ with the fermion current being $J^\lambda = \bar{\psi}_0\gamma^\lambda \psi_0\,.$ 
 Integrating over $B_{\mu\nu}$ we get 
\begin{align}
\label{Z_B}
\mathcal{Z}_B 
 \sim \, &\exp \frac{i}{4}\int d^4x d^4y\, \left(-Z^2 M_R^2 F_{\mu\nu} \frac{1}{\square+ M_R^2}F^{\mu\nu}- 4Z e_R M^2_R A_\mu \frac{1}{(\square+M^2_R)} J_\mu \right. \notag\\ 
& \qquad\qquad \left.\left.
+ 2 e_R^2 M_R^2 J_\lambda \frac{1}{\square(\square + M_R^2)}J^\lambda   
+ v^2  \Sigma^{\mu\nu}\frac{1}{\square+ M_R^2}\Sigma_{\mu\nu}\right.\right. \notag\\ 
& \qquad\qquad \left.-2v Z M_R\varepsilon^{\mu\nu\rho\lambda} A_\mu \frac{1}{\square + M^2_R}\partial_\nu \Sigma_{\rho\lambda}+ 2ve_R M_R \varepsilon^{\mu\nu\rho\lambda}\Sigma_{\mu\nu}\frac{1}{\Box(\square+M^2_R)} \partial_\rho J_\lambda\right) \,.
\end{align}
The first two terms and fifth term in the integral contribute to the action of $A_\mu$\,, which is now integrated over to get the force law. The integral over $A_\mu$ now reads
\begin{align}
\label{Z_A}
\mathcal{Z}_A = \int {\D}A_\mu &\exp i\int d^4x\left[-\frac{1}{4}F_{\mu\nu}\left(1 + \frac{Z^2 M_R^2}{\square+ M_R^2}\right) F^{\mu\nu}
-\frac{1}{2\xi}(\partial_\mu A^\mu)^2
\right. \notag\\ 
& \qquad\qquad \left. - A_\mu \left(e_RJ^\mu+\frac{e_R Z M^2_R}{\square + M^2_R}J^\mu + v ZM_R\frac{1}{\square + M^2_R}K^\mu\right) \right]\,, 
\end{align}
where we have written $K^\mu =\frac{1}{2} \varepsilon^{\mu\nu\rho\lambda}\partial_\nu \Sigma_{\rho\lambda}\,.$ Integration over $A_\mu$ produces
\begin{align}\label{ZA_J-J}
\mathcal{Z}_A \sim\,\exp\left(-\frac{i}{2}\int d^4x d^4y \tilde{J}^\mu(x) G_{\mu\nu}(x-y) 
 \tilde{J}^\nu(y) \right)\,,
\end{align} 
where $G_{\mu\nu}$ is the Green function calculated in Eq.~(\ref{Green-A}) and $\tilde{J}^\mu = \left(1+ \frac{Z M^2_R}{\square + M^2_R}\right)e_R J^\mu + v Z M_R\frac{1}{\square + M^2_R}K^\mu$\,.
To get the net interaction potential between fermions we now add the third term of Eq.~(\ref{Z_B})
to the above integral and convert the result to momentum space. Finally we  are left with the effective current-current interaction
\begin{equation}\label{Coulomb-action}
\frac{e_R^2}{2}\int \frac{d^4p}{(2\pi)^4} J^\mu(-p)\left(\frac{(1-Z)^2}{1+Z^2}\frac{1}{p^2-M_R^2(1+Z^2)} + \frac{2Z}{1+Z^2}\frac{1}{p^2} \right)J_\mu(p)\,.
\end{equation}
This represents, for the current of non-relativistic fermions, a Yukawa potential in the leading order along with a very small Coulomb correction. 
The expression \eqref{Z_A} also includes the  vortex-vortex and vortex-fermion interaction terms  and combining them with the relevant terms in $\mathcal{Z}_B$ we get 
\begin{equation}\label{vortex-vortex}
-\frac{v^2}{4}\int \frac{d^4p}{(2\pi)^4} \Sigma_{\mu\nu}(-p) \frac{1}{p^2 - M_R^2(1+Z^2)} \Sigma^{\mu\nu}(p)\,,
\end{equation}
which gives the interaction between two vortex lines and
\begin{equation}\label{vortex-fermion}
 \frac{iv e_R(1-Z) M_R}{2} \varepsilon^{\mu\nu\rho\lambda}\int \frac{d^4p}{(2\pi)^4} \Sigma_{\mu\nu}(-p) \frac{1}{p^2(p^2 - M_R^2(1+Z^2))}p_\rho J_\lambda(p)\,,
\end{equation}
which gives the vortex-fermion interaction. Since in the non-relativistic static limit we can write $\varepsilon^{0ijk}\partial_j\frac{1}{\square}J_k \sim S^i$, the spin magnetic moment density of a static electron, the above expression gives an effective vortex-spin interaction.
\section{Conclusion}
In this paper we analyzed the interaction of vortices in an Abelian Higgs model with charged fermions. This may be thought of as a field theoretic description of a type II superconductor with thin tubes of magnetic flux, in which unpaired electrons coexist with the charged pairs and interact electromagnetically through their minimal coupling with the photon. Dual formulation of the system using the four dimensional relativistic theory leads to a nonlocal interaction term between the antisymmetric tensor field and fermions, equivalent to a gauge field coupled to the spin density current of the fermions. This provides a post-facto justification of working with a relativistic formulation in four dimensions, for the spin of fermions appears naturally in it. 

One motivation of this work was to see if the vortex-electron interaction could give rise, in the dual picture, to the nonlocal coupling of the two-form field with electrons proposed earlier in~\cite{Choudhury:2015rua}. We found this, as an `emergent' interaction involving the spin current of the electrons that does not appear in the original way of writing the model but emerges in the process of dualization. Often the dual picture of the Abelian Higgs model in the context of a type II superconductor is studied as a nonrelativistic field theory (often in two spatial dimensions), leading to the disorder field~\cite{Kleinert:1982dz,Kleinert:2008zzb}, analogous to the antisymmetric tensor potential. We note however that since spin has to be introduced by hand in a non-relativistic theory, this interaction with the spin current would not have emerged from the non-relativistic field theory calculations usually done for superconductors. 

We have also found, as had been shown earlier, that this interaction generates a $B\wedge F$ term in one-loop effective action. This increases the mass of both gauge fields $A_\mu$ and $B_{\mu\nu}$\,, which should decrease the penetration depth. It was also shown earlier that the nonlocal interaction gives rise to a linear attractive potential between two non-relativistic fermions~\cite{Chatterjee:2016liu} and thus spatially localized fermion pairs would appear. We will investigate this phenomenon in future work. Here we have found something else -- a Coulomb potential between two charges, corresponding to a medium with a very high dielectric constant $\kappa\sim (2Z)^{-1}$\,. Thus 
we see from general principles that a material, otherwise a type II superconductor, will gain characteristics of a dielectric if unpaired electrons appear in it. 

We have also found the general forms of vortex-vortex and fermion-vortex interactions for this system. 
It should be possible to reduce our calculations to 2+1 dimensions and find the effective vortex-fermion, fermi-fermi, and vortex-vortex interactions in planar type II superconductors with unpaired electrons. It is also possible to consider temperature dependence of the coupling constants and investigate critical phenomena in the vortex-electron system in the dual picture presented in this paper. We leave these for future work.

\medskip

This research did not receive any specific grant from funding agencies in the public, commercial, or not-for-profit sectors.

\end{document}